\documentclass{ifacconf}
\usepackage{filecontents}

\makeatletter
\let\old@ssect\@ssect 
\makeatother
\usepackage{natbib}  
\usepackage{hyperref}
\makeatletter
\def\@ssect#1#2#3#4#5#6{%
	\NR@gettitle{#6}
	\old@ssect{#1}{#2}{#3}{#4}{#5}{#6}
}
\usepackage{graphicx}      
\usepackage{amsmath,amssymb,amsfonts}
\usepackage{algorithm}
\usepackage{algpseudocode}
\usepackage{graphicx}
\usepackage{textcomp}
\usepackage{xcolor}
\usepackage{stfloats}
\usepackage{booktabs}
\usepackage{enumitem}
\begin{document}
\begin{frontmatter}

\title{A Weighted Least-Squares Method for Non-Asymptotic Identification of Markov Parameters from Multiple Trajectories} 

\thanks[footnoteinfo]{This work was supported by VINNOVA Competence Center AdBIOPRO, contract [2016-05181] and by the Swedish Research Council through the research environment NewLEADS (New Directions in Learning Dynamical Systems), contract [2016-06079], and contract 2019-04956.}

\author{Jiabao He, Cristian R. Rojas and H\r{a}kan Hjalmarsson} 

\address{Division of Decision and Control Systems, School of Electrical Engineering and Computer Science, KTH Royal Institute of Technology, 100 44 Stockholm, Sweden(e-mail: jiabaoh, crro, hjalmars@kth.se).}

\begin{abstract}                
Markov parameters play a key role in system identification. There exists many algorithms where these parameters are estimated using least-squares in a first, pre-processing, step, including subspace identification and multi-step least-squares algorithms, such as Weighted Null-Space Fitting. Recently, there has been an increasing interest in non-asymptotic analysis of estimation algorithms. In this contribution we identify the Markov parameters using weighted least-squares and present non-asymptotic analysis for such estimator. To cover both stable and unstable systems, multiple trajectories are collected. We show that with the optimal weighting matrix, weighted least-squares gives a tighter error bound than ordinary least-squares for the case of non-uniformly distributed measurement errors. Moreover, as the optimal weighting matrix depends on the system's true parameters, we introduce two methods to consistently estimate the optimal weighting matrix, where the convergence rate of these estimates is also provided. Numerical experiments demonstrate improvements of weighted least-squares over ordinary least-squares in finite sample settings.
\end{abstract}

\begin{keyword}
Non-asymptotic identification,  Markov parameters, weighted least-squares.
\end{keyword}

\end{frontmatter}

\section{Introduction}  \label{Sct1}

There has been a recent resurgence of interest in identifying state-space models for dynamic systems. Solutions with asymptotic convergence guarantees have been well established in classical system identification, such as the celebrated Ho-Kalman algorithm \citep{Ho1966effective} and subspace identification methods \citep{Van2012subspace}. The focus of recent interest is non-asymptotic analysis, i.e., to provide finite sample probabilistic error bounds. The motivation for such analysis mainly arises in identification for different control tasks. For instance, the effectiveness of the control policy for the linear quadratic regulator with unknown dynamics hinges on the accuracy of identification, where a deeper understanding of the finite-sample behavior of estimation schemes constitutes an important objective \citep{jedra2022finite}. Also, in robust control problems, conducting such analyses proves instrumental in striking data-accuracy trade-offs \citep{oymak2021revisiting}. 

When the state can be directly measured, such linear time-invariant (LTI) systems are known as fully observed systems. Sharp finite-sample analysis of this type of systems using ordinary least-squares (OLS) is presented in \cite{sarkar2019near} and \cite{simchowitz2018learning}. Furthermore, for an uncontrolled system, a sample complexity lower bound is provided in \cite{jedra2022finite} and OLS is proven to be the optimal estimator. When the state is not available, such systems are known as partially observed systems. In this case, instead of recovering the systems matrices directly, the Markov parameters are often identified first, and subsequently used to determine a state-space realization. The Markov parameters have also been used to obtain end-to-end robust controllers \citep{tu2017non}. In terms of learning the Markov parameters, the most prevalent method is least-squares \citep{Sarkar2019FiniteTimeSI,tsiamis2019finite,oymak2019non,simchowitz2019learning,zheng2020non}. All these work suggest that the convergence rate is $\mathcal{O}(N^{-\frac{1}{2}})$, where $N$ is the number of samples. It should be mentioned that in order to cover both stable and unstable systems, multiple trajectories are usually required \citep{sun2020finite}, while for stable and marginally stable systems, a single trajectory is sufficient. The reason behind this is that an error state term becomes unbounded if the system is unstable, whereas this term can be replaced by the initial state in each trajectory that is zero by restarting the experiment \citep{zheng2020non}. Moreover, based on the fact that a Hankel matrix built from the Markov parameters usually has low rank, least-squares with low-rank regularization \citep{sun2020finite,djehiche2022efficient,fattahi2021learning} and low-rank approximation \citep{lee2022improved} are also used to estimate the Markov parameters.

Unlike fully observed systems, finding an optimal algorithm for partially observed systems is still an open problem. Hence, OLS continues to be extensively used for partially observed systems. However, it is usually not the best linear unbiased estimator (BLUE) as it ignores the unequal variance of the measurement errors. Instead, if the distribution of the noise is available, then the more general estimator, weighted least-squares (WLS) is known to be BLUE. Thus, a natural question is whether it is possible to achieve tighter finite sample error bounds using WLS as compared to OLS in the case of non-uniformly distributed measurement errors.

In this contribution, we use WLS to identify the Markov parameters of partially observed systems and provide a non-asymptotic analysis for such estimator. In order to cover both stable and unstable systems, multiple trajectories are collected. We show that with the optimal weighting matrix, WLS gives a tighter error bound than OLS. Although the convergence rate is still $\mathcal{O}(N^{-\frac{1}{2}})$, WLS has a smaller scaling factor. As far as we know, this is the first comparison between WLS and OLS in the non-asymptotic regime. Moreover, as the optimal weighting matrix depends on the system's true parameters, we introduce two methods to consistently estimate the optimal weighting matrix. Our analysis reveals that the gap between WLS with the optimal weighting matrix and the estimated weighting matrix decays with $\mathcal{O}(N^{-1})$. Thus, as the sample size increases, WLS with the estimated weighting matrix approaches the optimal one quickly. 

The disposition of the paper is as follows: The problem is formulated in Section \ref{Sct2}. The non-asymptotic analysis of WLS with the optimal weighting matrix is provided in Section \ref{Sct3}. In Section \ref{Sct4}, we show how to estimate the optimal weighting matrix. Two examples are provided in Section \ref{Sct5} to demonstrate
improvements of WLS over OLS. Finally, conclusions are provided in Section \ref{Sct6}.

Notations: For a matrix $X$ with appropriate dimensions, $X^\top$, $X^{-1}$, $\lVert X \rVert$, $\lVert X \rVert_F$ and $\lambda_{\rm{min}}(X)$ denote its transpose, inverse, spectral norm, Frobenius norm and minimum eigenvalue, respectively. $X \succcurlyeq 0$ means that $X$ is positive semi-definite. The matrix $I_N$ is the identity matrix with  dimension ${N\times N}$. The notation $X \otimes Y$ is the Kronecker product between matrices $X$ and $Y$. The multivariate normal distribution with mean $\mu$ and covariance $\Sigma$ is denoted as $\mathcal{N}(\mu,\Sigma)$.

\section{Problem Statement} \label{Sct2}

Consider the following discrete-time LTI system on innovations form:
\begin{subequations} \label{E1}
	\begin{align}
		x_{t + 1} &= Ax_{t}  + Bu_{t} + Ke_{t}, \label{E1a}\\
		y_{t} &= Cx_{t} + Du_{t} + e_{t}, \label{E1b}		
	\end{align}
\end{subequations}
where $x_{t}\in \mathbb{R}^{n_x}$, $u_{t}\in \mathbb{R}^{n_u}$, $y_{t}\in \mathbb{R}^{n_y}$ and $e_{t}\in \mathbb{R}^{n_y}$ are the state, input, output and innovation, respectively.  We assume that $u_{t}$ and $e_{t}$ are independent and identically distributed Gaussian random variables, i.e., $u_{t} \sim \mathcal{N}(0,\sigma_u^2)$ and $e_{t} \sim \mathcal{N}(0,\sigma_e^2)$, respectively. In addition, letting the initial state $x_{0} = 0$, the output $y_{t}$ can be expanded recursively as 
\begin{equation} \label{E2}
	{y_t} =  \sum\limits_{k = 0}^{t-1}  CA^k(B{u_{t-k-1}} + K{e_{t-k-1}})  + D{u_t}  + {e_t}.
\end{equation}

To allow for any spectral radius of $A$, i.e., to deal with a possibly unbounded state $CA^{T-1}x_{t-T+1}$, we similarly collect data from multiple trajectories. To be specific, we have $N$ independent rollouts, and the system is excited for $T$ steps in each rollout. The resulting dataset is
\begin{equation} \label{E3}
	\left\{\left(y_t^{(i)},u_t^{(i)}\right): 1 \leq i \leq N, 0 \leq t \leq T-1\right\},
\end{equation}
where $t$ indexes the instant and $i$ denotes the rollout. To ease notation, for each rollout $i$, we organize the input, output and noise as
\begin{subequations} \label{E4}
	\begin{align}
		Y^{(i)} &= \begin{bmatrix}y_0^{(i)} & y_1^{(i)} & y_2^{(i)} & \cdots & y_{T-1}^{(i)} \end{bmatrix}, \label{E4a}\\
		U^{(i)} &= \begin{bmatrix}
			u_0^{(i)}&u_1^{(i)}&u_2^{(i)}& \cdots &u_{T-1}^{(i)}\\
			0&u_0^{(i)}&u_1^{(i)}&\cdots &u_{T-2}^{(i)}\\
			0&0&u_0^{(i)}&\cdots &u_{T-3}^{(i)}\\
			\vdots & \vdots & \vdots &\ddots & \vdots \\
			0&0&0& \cdots &u_0^{(i)}
		\end{bmatrix}, \label{E4b}
	\end{align}
\end{subequations}
and $E^{(i)}$ has the same structure as $U^{(i)}$. In this way, 
the output trajectory of each rollout can be written as
\begin{equation} \label{E5}
	Y^{(i)} = GU^{(i)} + HE^{(i)},
\end{equation}
where 
\begin{subequations} \label{E6}
	\begin{align}
		G &= \begin{bmatrix}
			D&CB&CAB& \cdots &CA^{T-2}B
		\end{bmatrix}, \label{E6a}\\
		H &= \begin{bmatrix}
			I&CK&CAK& \cdots &CA^{T-2}K
		\end{bmatrix} \label{E6b}		
	\end{align}
\end{subequations}
are the Markov parameters with respect to the input and innovations, respectively.
Furthermore, using all rollouts, we define 
\begin{subequations} \label{E7}
	\begin{align}
		Y&= \begin{bmatrix}
			Y^{(1)}&Y^{(2)}&Y^{(3)}& \cdots &Y^{(N)}
		\end{bmatrix},\\
		U&= \begin{bmatrix}
			U^{(1)}&U^{(2)}&U^{(3)}& \cdots &U^{(N)}
		\end{bmatrix},\\
		E&= \begin{bmatrix}
			E^{(1)}&E^{(2)}&E^{(3)}& \cdots &E^{(N)}
		\end{bmatrix}.
	\end{align}
\end{subequations}
According to \eqref{E5} and \eqref{E7}, the output of all rollouts is 
\begin{equation} \label{E8}
	Y = GU + HE.
\end{equation}
Next, we explain how to estimate Markov parameters $G$ using OLS and WLS based on \eqref{E8}.

\section{Identifying Markov Parameters} \label{Sct3}

\subsection{Ordinary Least-Squares} \label{Sct3.1}

A prevalent estimator for estimating Markov parameters is OLS. The OLS estimator of $G$ is given by
\begin{equation} \label{E9}
	\hat G_{\rm{ols}}= {\rm{arg}} \mathop{\rm {min}} \limits_{X} \sum\limits_{i = 1}^N \lVert Y^{(i)} - XU^{(i)}\rVert^2,
\end{equation}
or, equivalently, 
\begin{equation} \label{E10}
	\hat G_{\rm{ols}}= {\rm{arg}} \mathop{\rm {min}} \limits_{X} \lVert Y - XU\rVert_F^2. 
\end{equation}
The closed-form solution for the above problem is 
\begin{equation} \label{E11}
	\hat G_{\rm{ols}}=YU^\top(UU^\top)^{-1},
\end{equation}
and the estimation error is given by
\begin{equation} \label{E12}
	\tilde G_{\rm{ols}} := \hat G_{\rm{ols}} - G = HEU^\top(UU^\top)^{-1}.
\end{equation}
We have the following lemma regarding the non-asymptotic analysis of OLS:
\begin{lem} \label{Lma1}
	For any $0<\delta<1$, if the number of rollouts satisfies $N\geq N_{\rm{ols}}$, then, with probability at least $1-\delta$,
	\begin{equation} \label{E13}
		\lVert \tilde G_{\rm{ols}} \rVert \leq \frac {\sigma_e}{\sigma_u} C_{\rm{ols}}\sqrt{\frac {1}{N}},
	\end{equation}
	where 
	\begin{equation} \label{E14}
		\begin{split}
			N_{\rm{ols}}&=8n_uT+4(n_u+n_y+4){\rm{log}}(2T/\delta), \\
			C_{\rm{ols}}&= 16\lVert H \rVert \sqrt{ \left(\frac {2T^3+3T^2+T}{3}\right)\left(n_u+n_y\right)
				{\rm{log}}\left(\frac{18T}{\delta}\right)}.
		\end{split}
	\end{equation}
\end{lem}
\begin{pf}
	After replacing $B_w$, $D_v$, $w_{t}$ and $v_{t}$ in the nominal state-space form used in \cite{zheng2020non} with $K$, $I$ and $e_{t}$ in the innovations form \eqref{E1}, this lemma can be easily derived from Theorem 1 in \cite{zheng2020non}. \hfill  $\blacksquare$
\end{pf}
\vspace{-1.5mm}
\subsection{Weighted Least-Squares} \label{Sct3.2}
\vspace{-1.5mm}
OLS gives a consistent estimate mainly due to the zero mean of the noise term $HE$. However, other statistical properties of the noise are not fully used. In particular, OLS does not benefit from the correlation structure of the noise term. Assuming that this structure is known, WLS is BLUE with superior statistical properties over OLS. For simplicity of illustration, we consider the SISO case in this section, and will show that our method can be applied to MIMO cases in numerical examples in Section \ref{Sct5}.

First, to reveal the distribution of $HE$, we express it as 
\begin{equation} \label{E15}
	HE= \mathcal{E}\mathcal{K}_H,
\end{equation}
where 
\begin{subequations} \label{E16}
	\begin{align}
		\mathcal{E} &= \begin{bmatrix}
			\mathcal{E} ^{(1)}&\mathcal{E} ^{(2)}&\mathcal{E} ^{(3)}& \cdots &\mathcal{E} ^{(N)}
		\end{bmatrix}, \label{E16a}\\
		\mathcal{E} ^{(i)} &= \begin{bmatrix}
			e_0^{(i)}&e_1^{(i)}&e_2^{(i)}& \cdots &e_{T-1}^{(i)}\end{bmatrix}, \label{E16b}\\
		\mathcal{K}_H &= I_N \otimes \mathcal{T}_H, \label{E16c}\\
		\mathcal{T}_H &= \begin{bmatrix}
			1&CK&CAK& \cdots &CA^{T-2}K\\
			0&1&CK&\cdots &CA^{T-3}K\\
			0&0&1&\cdots &CA^{T-4}K\\
			\vdots & \vdots & \vdots &\ddots & \vdots \\
			0&0&0& \cdots &1
		\end{bmatrix}.  \label{E16d}
	\end{align}
\end{subequations}

Compared to the left side of \eqref{E15}, the main benefit of the right side is that the variance of $\mathcal{E}\mathcal{K}_H$ can be easily obtained as ${{\rm{Var}}(\mathcal{E}\mathcal{K}_H)} = \sigma_e^2\mathcal{K}_H^\top \mathcal{K}_H$. In this way, with the optimal weighting matrix
\begin{equation} \label{E17}
    W^{*} = (\mathcal{K}_H^\top \mathcal{K}_H)^{-1} = I_N \otimes (\mathcal{T}_H^\top \mathcal{T}_H)^{-1},
\end{equation}
$G$ can be estimated using WLS based on the model \eqref{E8}, which gives 
\begin{equation} \label{E18}
	\hat G_{\rm{wls}}^{*} = YW^{*}U^\top(UW^{*}U^\top)^{-1}.
\end{equation}
Furthermore, the estimation error is
\begin{equation} \label{E19}
	\tilde G_{\rm{wls}}^{*} := \hat G_{\rm{wls}}^{*}  - G = \mathcal{E}\mathcal{K}_H^{-\top}U^{\top}(UW^{*}U^\top)^{-1}.
\end{equation} 

Before presenting our main contribution, namely a non-asymptotic error analysis of WLS, we compare the variance of the estimation errors between OLS and WLS, which verifies that WLS has a smaller variance. The variance matrices of OLS and WLS are 
\begin{subequations} \label{E20}
	\begin{align}
		{\rm{Var}}(\tilde G_{\rm{ols}}) &:=  \sigma_e^2(UU^\top)^{-1}U(W^{*})^{-1}U^\top(UU^\top)^{-1},\\
		{\rm{Var}}(\tilde G_{\rm{wls}}^{*} ) &:=  \sigma_e^2(UW^{*}U^\top)^{-1},
	\end{align}
\end{subequations}
respectively. In this way,
\begin{equation} \label{E21}
	{\rm{Var}}(\tilde G_{\rm{ols}}) - {\rm{Var}}(\tilde G_{\rm{wls}}^{*} ) = \sigma_e^2(UU^\top)^{-1}\Gamma(UU^\top)^{-1},
\end{equation}
where $\Gamma = U(W^{*})^{-1}U^\top-UU^\top(UW^{*}U^\top)^{-1}UU^\top$. 
According to the Shur complement, $\Gamma \succcurlyeq 0$ is equivalent to 
\begin{equation} \label{E22}
	\begin{split}
		&\begin{bmatrix}
		U(W^{*})^{-1}U^\top &UU^\top\\
		UU^\top&UW^{*}U^\top
	\end{bmatrix} = \\
&	\begin{bmatrix}
		U&0\\
		0&U
	\end{bmatrix}\begin{bmatrix}
		(W^{*})^{-1} &I\\
		I&W^{*}
	\end{bmatrix}\begin{bmatrix}
		U^\top &0\\
		0&U^\top
	\end{bmatrix}\succcurlyeq 0.
	\end{split}
\end{equation}
Since $W^{*} \succcurlyeq 0$, we have $\begin{bmatrix}
	{W^{*}}^{-1} &I\\
	I&W^{*}
\end{bmatrix} \succcurlyeq 0$. As a result, ${\rm{Var}}(\tilde G_{\rm{ols}}) - {\rm{Var}}(\tilde G_{\rm{wls}}^{*} ) \succcurlyeq 0$, which means that WLS with the optimal weighting matrix is better than OLS. Now, we present the non-asymptotic analysis of WLS.

\begin{thm} \label{Th1}
	For any $0<\delta<1$, if the number of rollouts satisfies $N\geq N_{\rm{wls}}$, then,  with probability at least $1-\delta$,
	\begin{equation} \label{E23}
		\lVert \tilde G_{\rm{wls}}^{*} \rVert \leq \frac {\sigma_e}{\sigma_u} C_{\rm{wls}}^{*}\sqrt{\frac {1}{N}},
	\end{equation}
	where 
	\begin{equation} \label{E24}
		\begin{split}
			N_{\rm{wls}} &=8n_uT+2(n_u+n_y+8){\rm{log}}(2T/\delta), \\
			C_{\rm{wls}}^{*} &= 16\lVert H \rVert \sqrt{ \left(\frac{T^3+T^2}{2}\right)\left(n_u+n_y\right)
				{\rm{log}}\left(\frac{18T}{\delta}\right)}.
		\end{split}
	\end{equation}
\end{thm}

\begin{pf}
	The main idea is to simplify the estimation error of WLS in \eqref{E19} by utilizing the structure of $U$ and $\mathcal{T}_H$. First, the term $UW^{*}U^\top$ in \eqref{E19} can be expressed as
	\begin{equation} \label{E25}
		\begin{split}
			&UW^{*}U^\top = U((I_N \otimes \mathcal{T}_H)^{\top}(I_N \otimes \mathcal{T}_H))^{-1}U^\top  \\ 
			= &\sum\limits_{i = 1}^N{U^{(i)}\mathcal{T}_H^{-1}\mathcal{T}_H^{-\top}U^{(i)\top}} 
			= \sum\limits_{i = 1}^N{\mathcal{T}_H^{-1}U^{(i)}U^{(i)\top}\mathcal{T}_H^{-\top}} \\
			= &\mathcal{T}_H^{-1}UU^{\top}\mathcal{T}_H^{-\top}, 
		\end{split}
	\end{equation}
	where the third equality is based on the fact that both $U^{(i)}$ and $\mathcal{T}_H^{-1}$ are upper-triangular Toeplitz matrices, which makes their multiplication commutative. Similarly, the term $\mathcal{E}\mathcal{K}_H^{-\top}U^{\top}$ in \eqref{E19} can be expressed as 
	\begin{equation} \label{E26}
	\begin{split}
		&\mathcal{E}\mathcal{K}_H^{-\top}U^{\top} = \mathcal{E}(I_N \otimes \mathcal{T}_H)^{-\top}U^{\top} 	= \mathcal{E}(I_N \otimes \mathcal{T}_H^{-\top})U^{\top} \\
		= &\mathcal{E}\begin{bmatrix}
			\mathcal{T}_H^{-\top}U^{(1)\top}\\
			\vdots\\
			\mathcal{T}_H^{-\top}U^{(N)\top}\\
		\end{bmatrix}= \mathcal{E}\begin{bmatrix}
		U^{(1)\top}\mathcal{T}_H^{-\top}\\
		\vdots\\
		U^{(N)\top}\mathcal{T}_H^{-\top}\\
    	\end{bmatrix} = \mathcal{E}U^{\top}\mathcal{T}_H^{-\top}.
	\end{split}
	\end{equation}
    After multiplying \eqref{E25} and \eqref{E26} together, \eqref{E19} becomes 
	\begin{equation} \label{E27}
		\begin{split}
		\tilde G_{\rm{wls}}^{*} &= \mathcal{E}U^{\top}\mathcal{T}_H^{-\top}(\mathcal{T}_H^{-1}UU^{\top}\mathcal{T}_H^{-\top})^{-1} \\
		&=\mathcal{E}U^{\top}(UU^{\top})^{-1}\mathcal{T}_H. 
	     \end{split}
	\end{equation}
	Now $\tilde G_{\rm{wls}}^{*}$ contains two multiplications of random matrices, $\mathcal{E}U^{\top}$ and $UU^{\top}$, and we need the following lemmas in \cite{zheng2020non} to deal with those multiplications.
	\begin{lem} \label{Lma2}
		For a fixed $0<\delta<1$, if $N\geq 8n_uT+16{\rm{log}}(T/\delta)$, we have with probability at least $1-\delta$ that
		\begin{equation} \label{E28}
			\lambda_{\rm{min}}(UU^{\top})\geq \frac{1}{4}\sigma_u^2N.
		\end{equation}
	\end{lem}
	\begin{lem} \label{Lma3}
		For a fixed $0<\delta<1$, if $N\geq 2(n_u+n_y){\rm{log}}(T/\delta)$, we have with probability at least $1-\delta$ that
		\begin{equation} \label{E29}
			\lVert \mathcal{E}U^\top \rVert \leq 2\sigma_e\sigma_u\sqrt{2T(T+1)N(n_u+n_y)log(9T/\delta)}.
		\end{equation}
	\end{lem}
	According to the above lemmas, we conclude that for any $0<\delta<1$, if the number of rollouts satisfies $N\geq 8n_uT+2(n_u+n_y+8){\rm{log}}(2T/\delta)$, we have with probability at least $1-\delta$ that
	\begin{equation*}
		\lVert \tilde G_{\rm{wls}}^{*}\rVert \leq \frac
		{\sigma_e}{\sigma_u} 8\lVert \mathcal{T}_H \rVert \sqrt{ 2T(T+1)\left(n_u+n_y\right)
			{\rm{log}}\left(18T/\delta\right)}\sqrt{\frac {1}{N}}.
	\end{equation*}
	To directly compare the error bounds of OLS and WLS, we need to deal with $\mathcal{T}_H$ which is an upper-triangular Toeplitz matrix based on $H$. According to the structure of $\mathcal{T}_H$ and $H$, we have 
	\begin{equation} \label{E30}
		\lVert \mathcal{T}_H \rVert \leq \lVert \mathcal{T}_H \rVert_F \leq \sqrt{T}\lVert H \rVert.
	\end{equation}
	After replacing $\lVert \mathcal{T}_H \rVert$ with the larger bound $\sqrt{T}\lVert H \rVert$, and defining $C_{\rm{wls}}^{*}$ as in \eqref{E24}, the proof is completed. \hfill  $\blacksquare$
\end{pf}

\begin{rem} \label{rem1}
	Comparing Lemma \ref{Lma1} and Theorem \ref{Th1}, we can see WLS and OLS both have the convergence rate of $\mathcal{O}(N^{-\frac{1}{2}})$. However, the scaling factors satisfy $C_{\rm{wls}}^{*} < C_{\rm{ols}}$, and the required minimal number of rollouts $N_{\rm{wls}} < N_{\rm{ols}}$. These comparisons suggest that for a fixed probability $1-\delta$, WLS attains a tighter error bound than OLS.
\end{rem}

\begin{rem} \label{rem2}
	Similar to Lemma \ref{Lma1}, Theorem \ref{Th1} is not tight with respect to system dimensions. First, there is a redundancy in the length $T$, in the sense that there are $n_yn_uT$ parameters in $G$ to be estimated, while Lemma \ref{Lma1} and Theorem \ref{Th1} state that $\mathcal{O}(n_uT)$ trajectories are required, where each of them provides $n_yT$ measurements. Second, both constants $C_{\rm{ols}}$ in Lemma \ref{Lma1} and $C_{\rm{wls}}^{*}$ in Theorem \ref{Th1} depend polynomially on the length $T$, indicating that more Markov parameters require more trajectories to achieve the same accuracy. Meanwhile, as $T$ is also the length of each experiment, a larger $T$ means that we have more data for identification. Generally speaking, it is not straightforward how to analyze the impact of $T$, as its value may affect differently the estimation error for SISO and MIMO systems, which is further elaborated in Section \ref{Sct5}. More discussions about this can be found in \cite{zheng2020non}.
\end{rem}
\vspace{-1.5mm}
\section{Estimating the Weighting Matrix} \label{Sct4}
\vspace{-1.5mm}
Since the optimal weighting matrix $W^{*}$ depends on the system's true Markov parameters $H$, i.e., $\left\{CA^{i}K\right\}$, we now show how to consistently estimate $H$. First, we introduce the following predictor form of system \eqref{E1}:
\begin{subequations} \label{E31}
	\begin{align}
		x_{t + 1} &= A_Kx_{t}  + B_Ku_{t} + Ky_{t}, \label{E2a}\\
		y_{t} &= Cx_{t} + Du_{t} + e_{t}, \label{E2b}		
	\end{align}
\end{subequations}
where $A_K = A-KC$ and $B_K =  B-KD$.
Similar to \eqref{E2}, the output of each rollout can be expanded recursively as 
\begin{equation} \label{E32}
	y_t^{(i)} =  \sum\limits_{k = 0}^{t-1}  \left(CA_K^kB_K{u_{t-k-1}^{(i)}}  + CA_K^kK{y_{t-k-1}^{(i)}}\right)  + Du_t^{(i)} + e_t^{(i)}.
\end{equation}
Letting $t=T-1$, we have
\begin{equation} \label{E33}
	y_{T-1}^{(i)} = G_Ku^{(i)} + H_Ky^{(i)} + e_{T-1}^{(i)},
\end{equation}
where
\begin{subequations} \label{E34}
	\begin{align}
		G_K &= \begin{bmatrix}
			CA_K^{T-2}B_K& \cdots &CA_KB_K&CB_K&D
		\end{bmatrix}, \\
		H_K &= \begin{bmatrix}
			CA_K^{T-2}K& \cdots &CA_KK&CK&0
		\end{bmatrix}, \\
		u^{(i)} &= \begin{bmatrix}
		(u_0^{(i)})^\top&\cdots&(u_{T-3}^{(i)})^\top&(u_{T-2}^{(i)})^\top&(u_{T-1}^{(i)})^\top
		\end{bmatrix}^\top, \\
		y^{(i)} &= \begin{bmatrix}
			(y_0^{(i)})^\top&
			\cdots &
			(y_{T-3}^{(i)})^\top&
			(y_{T-2}^{(i)})^\top&
			(y_{T-1}^{(i)})^\top
		\end{bmatrix}^\top.
	\end{align}
\end{subequations}
Combining all rollouts, we have
\begin{equation} \label{E35}
	y_{T-1} = G_Ku + H_Ky + e_{T-1},
\end{equation}
where
\begin{subequations} \label{E36}
	\begin{align}
		y_{T-1} &= \begin{bmatrix}
			y_{T-1}^{(1)}&y_{T-1}^{(2)}&y_{T-1}^{(3)}& \cdots &y_{T-1}^{(N)}
		\end{bmatrix}, \\
		e_{T-1} &= \begin{bmatrix}
			e_{T-1}^{(1)}&e_{T-1}^{(2)}&e_{T-1}^{(3)}& \cdots &e_{T-1}^{(N)}
		\end{bmatrix}, \\
		u &= \begin{bmatrix}
			u^{(1)}&u^{(2)}&u^{(3)}& \cdots &u^{(N)}
		\end{bmatrix}, \\
		y &= \begin{bmatrix}
			y^{(1)}&y^{(2)}&y^{(3)}& \cdots &y^{(N)}
		\end{bmatrix}.
	\end{align}
\end{subequations}
Now, the estimates of $G_K$ and $H_K$ given by OLS are
\begin{equation} \label{E37}
	{\begin{bmatrix}\hat G_K&\hat H_K\end{bmatrix}} = y_{T-1}{\begin{bmatrix}u\\y\end{bmatrix}}^\top{\left(\begin{bmatrix}u\\y\end{bmatrix}{\begin{bmatrix}u\\y\end{bmatrix}}^\top\right)}^{-1}.
\end{equation}
To date, we have consistent estimates of $G_K$ and $H_K$. Actually, to obtain the optimal weighting matrix $W^{*}$ that depends on $\left\{CA^iK\right\}$, only $\left\{CA_K^iK\right\}$ in $H_K$ are needed. Now we provide two methods to extract $\left\{CA^iK\right\}$ from $\left\{CA_K^iK\right\}$. It should be mentioned that, to ease notation, we write $\left\{CA^iK\right\}$ and $\left\{CA_K^iK\right\}$, while in practice we use their corresponding estimates.
\vspace{-1.5mm}
\subsection{Recursive Algorithm} \label{Sct4.1}
\vspace{-1.5mm}
The first method is quite straightforward. It is based on the observation that after we expand the multiplicity of $CA_K^iK$, $CA^iK$ can be determined by $CA_K^iK$ and all previous values of $CA^jK$, where $j<i$. To be specific, the following recursive algorithm is used \citep{Juang1993identification,Zhao2014subspace}: 
\begin{equation} \label{E27}
	CA^{i-1}K = CA_K^{i-1}K + \sum_{j=1}^{i-1}{(CA_K^{j-1}K) (CA^{i-j-1}K)}.
\end{equation} 
\vspace{-1.5mm}
\subsection{Ho-Kalman Algorithm}  \label{Sct4.2}
\vspace{-1.5mm}
Another method is based on the celebrated Ho-Kalman algorithm \citep{Ho1966effective}. As $H_K$ is the Markov parameter of the predictor form \eqref{E31}, to recover $\left\{CA^iK\right\}$, an alternative way is to obtain a realization of the matrices $\left\{C, A_K, K\right\}$ from $H_K$, and then $\left\{C, A, K\right\}$ can be obtained. It has been proven that as long as  $\left\{\widehat{CA_K^iK}\right\}$ is a consistent estimate of $\left\{CA_K^iK\right\}$, then using the Ho-Kalman algorithm, the realization of $\left\{C, A_K, K\right\}$ is consistent up to a similarity transformation \cite{oymak2021revisiting}. In this way, $\left\{\widehat{CA^iK}\right\}$ will be a consistent estimate of $\left\{CA^iK\right\}$. However, the accuracy of this algorithm is slightly worse than the recursive computing method, as the singular value decomposition (SVD) in the Ho-Kalman algorithm truncates the smaller singular values. We will see such a difference in the simulation part in Section \ref{Sct5}.
\vspace{-1.5mm}
\subsection{Convergence Rate}  \label{Sct4.3}
\vspace{-1.5mm}
Now we analyze the convergence rate of the estimated weighting matrix $\hat W$, as well as its performance compared to the optimal one. Using the expression for the inverse of a block matrix \citep{bernstein2009matrix}, $\hat H_K$ in \eqref{E37} is given as
\begin{equation} \label{E39}
	\hat H_K = y_{T-1}\Pi_{u}^{\perp}y^\top(y\Pi_{u}^{\perp}y^\top)^{-1},
\end{equation}
where 
$\Pi_{u}^{\perp} = I-u^\top(uu^\top)^{-1}u$, and the error is 
\begin{equation} \label{E40}
	\tilde H_K := \hat H_K - H_K = e_{T-1}\Pi_{u}^{\perp}y^\top(y\Pi_{u}^{\perp}y^\top)^{-1}.
\end{equation}
We then have the following lemma regarding the convergence rate of $\tilde H_K$.
\begin{lem} \label{cor1}
	The convergence rate of the error $\tilde H_K$ scales $\mathcal{O}(N^{-\frac{1}{2}})$.
\end{lem}
\begin{pf}
	Based on the innovation form \eqref{E1}, $y^{(i)}$ in \eqref{E34} can be written as
	\begin{equation} \label{E41}
		y^{(i)} = \Phi u^{(i)} + \Psi e^{(i)},
	\end{equation}
	where 
	\begin{equation} \label{E42a}
	\Phi = \begin{bmatrix}
		D&0&0& \cdots &0\\
		CB&D&0&\cdots &0\\
		CAB&CB&D&\cdots &0\\
		\vdots & \vdots & \vdots &\ddots & \vdots \\
		CA^{T-2}B&CA^{T-3}B&CA^{T-4}B& \cdots &D
	\end{bmatrix},
		\nonumber
\end{equation}
\begin{equation} \label{E42b}
	\Psi = \begin{bmatrix}
		I&0&0& \cdots &0\\
		CK&I&0&\cdots &0\\
		CAB&CK&I&\cdots &0\\
		\vdots & \vdots & \vdots &\ddots & \vdots \\
		CA^{T-2}K&CA^{T-3}K&CA^{T-4}K& \cdots &I
	\end{bmatrix},
	\nonumber
\end{equation}
and $e^{(i)}$ has the same structure as $u^{(i)}$ in \eqref{E34}. Thus, combining all rollouts, we have
	\begin{equation} \label{E43}
		y = \Phi u+ \Psi e.
	\end{equation}
After substituting \eqref{E43} into \eqref{E40}, we have
	\begin{equation} \label{E44}
			\tilde H_K = e_{T-1}\Pi_{u}^{\perp}(\Psi e)^\top \left((\Psi e)\Pi_{u}^{\perp}(\Psi e)^\top\right)^{-1}.
	\end{equation}
The eigendecomposition of the projection matrix $\Pi_{u}^{\perp}$ can be written as $\Pi_{u}^{\perp} = Q\Lambda Q^\top = Q_{n_1} Q_{n_1}^\top$, where $\Lambda$ is a diagonal matrix whose diagonal elements are 1 or 0, $Q$ is an orthogonal matrix, $Q_{n_1}$ is the first $n_1$ columns of $Q$, and $n_1$ is the number of 1's in $\Lambda$. In this way, \eqref{E44} can be rewritten as
	\begin{equation} \label{E44a}
		\tilde H_K = e_{T-1}Q_{n_1}(\Psi eQ_{n_1})^\top \left((\Psi eQ_{n_1})(\Psi eQ_{n_1})^\top\right)^{-1}.
	\end{equation}
Observe that $e_{T-1}Q_{n_1}$ and $eQ_{n_1}$ are independent and identical Gaussian random matrices, so according to Lemmas \ref{Lma2} and \ref{Lma3} we can similarly conclude that the convergence rate of $\tilde H_K$ is $\mathcal{O}(N^{-\frac{1}{2}})$. Furthermore, each block $\left\{\widetilde{CA_K^iK}\right\}$ in $\tilde H_K$ has the convergence rate $\mathcal{O}(N^{-\frac{1}{2}})$.  \hfill  $\blacksquare$
\end{pf}

Using the algorithms in Sections \ref{Sct4.1} and \ref{Sct4.2} to extract $\left\{\widehat{CA^iK}\right\}$ from $\left\{\widehat{CA_K^iK}\right\}$, we have the following lemma regarding the convergence rate of the estimation error $\left\{\widetilde{CA^iK}\right\}$, and the resulting error in the upper-Toeplitz matrix $\tilde{\mathcal{T}}_H :=\hat{\mathcal{T}}_H - \mathcal{T}_H$.

\begin{lem} \label{cor2}
	With the extraction algorithms in Sections \ref{Sct4.1} and \ref{Sct4.2}, the convergence rate of $\left\{\widetilde{CA^iK}\right\}$ scales  $\mathcal{O}(N^{-\frac{1}{2}})$, and the resulting error in $\tilde{\mathcal{T}}_H$ also scales $\mathcal{O}(N^{-\frac{1}{2}})$.
\end{lem}
\begin{pf}
	To conserve space, the complete proof is omitted here. For the recursive algorithm, its convergence rate is not hard to prove. For the convergence rate of the Ho-Kalman algorithm, we refer the readers to \cite{oymak2021revisiting}.  \hfill  $\blacksquare$
\end{pf}

Now we analyze the performance of WLS with the estimated weighting matrix $\hat W$. Similar techniques have been presented in the asymptotic regime \citep{Carroll1988effect,Chen1993iterative,Galrinho2018parametric}.  After replacing the optimal weighting matrix $W^{*}$ with its estimate $\hat W$ in \eqref{E18}, the estimation error  \eqref{E27} becomes 
\begin{equation} \label{E45}
	\begin{split}
		& \hat G_{\rm{wls}} - G = \mathcal{E}U^{\top}(UU^{\top})^{-1}\hat{\mathcal{T}}_H \\
		= & \mathcal{E}U^{\top}(UU^{\top})^{-1}\mathcal{T}_H + \mathcal{E}U^{\top}(UU^{\top})^{-1}(\hat{\mathcal{T}}_H - \mathcal{T}_H) \\
		= & \hat G_{\rm{wls}}^{*} - G + \mathcal{E}U^{\top}(UU^{\top})^{-1}(\hat{\mathcal{T}}_H - \mathcal{T}_H).
	\end{split}
\end{equation}
As we can see, since $\hat G_{\rm{wls}}^{*} - G$, $\mathcal{E}U^{\top}(UU^{\top})^{-1}$ and $(\hat{\mathcal{T}}_H - \mathcal{T}_H)$ both have convergence rate  $\mathcal{O}(N^{-\frac{1}{2}})$, the whole rate of the second term $\mathcal{E}U^{\top}(UU^{\top})^{-1}(\hat{\mathcal{T}}_H - \mathcal{T}_H)$ is $\mathcal{O}(N^{-1})$, which decays faster than the rate of $\hat G_{\rm{wls}}^{*} - G$. It means that the error caused by estimating the weighting matrix is dominated by the statistical error $\hat G_{\rm{wls}}^{*} - G$. In other words, WLS with the estimated weighting matrix $\hat W$ approaches the optimal weighting matrix $W^{*}$ closely, and outperforms OLS. Such behaviour is also observed in our numerical examples in Section \ref{Sct5}.

\vspace{-1.5mm}
\section{Simulation Examples} \label{Sct5}
\begin{figure}
	\centering
	\includegraphics[scale=0.19]{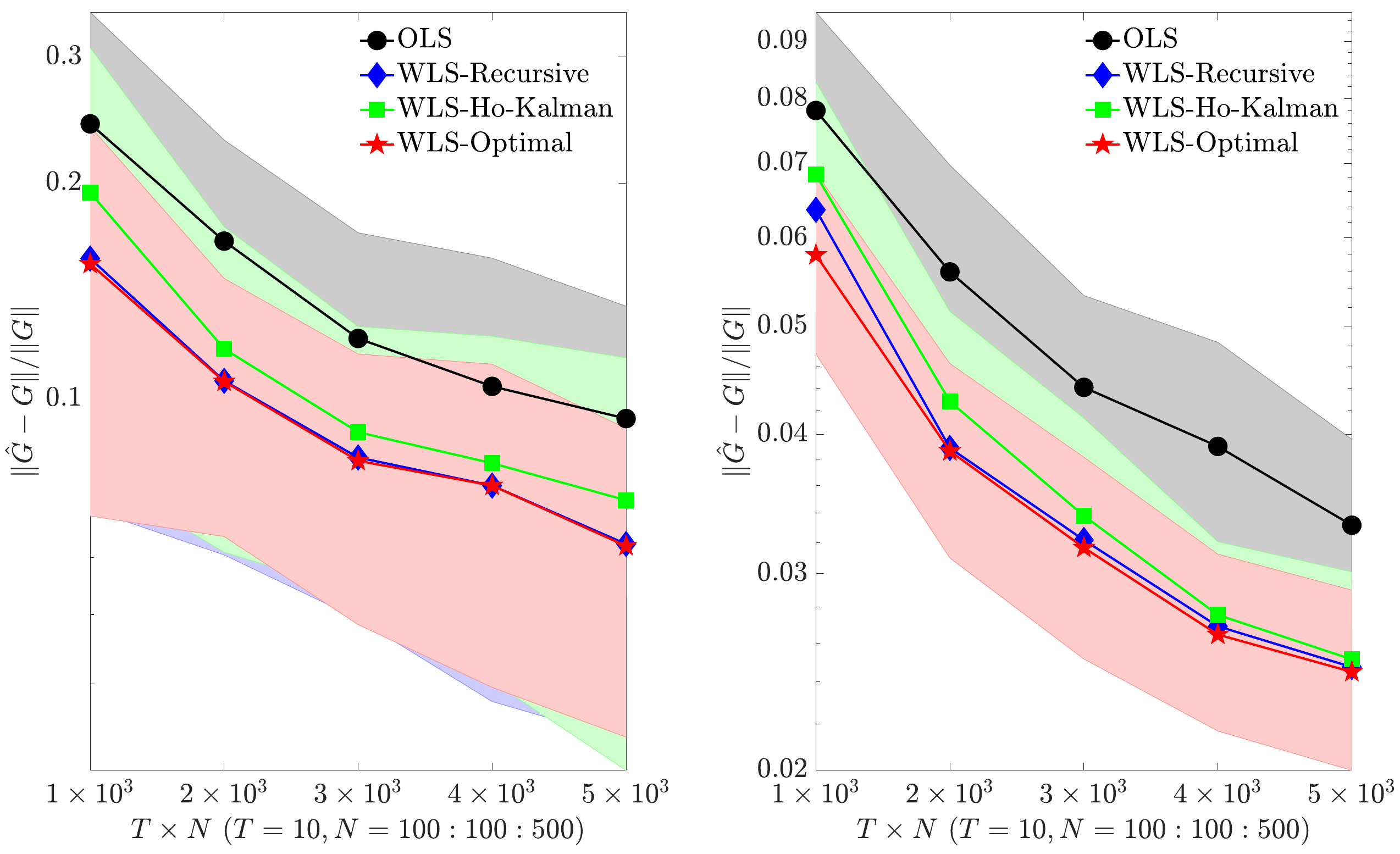}
	\caption{Estimation errors of SISO (left) and MIMO (right).}
	\label{F1}
\end{figure}

\begin{figure}
	\centering
	\includegraphics[scale=0.19]{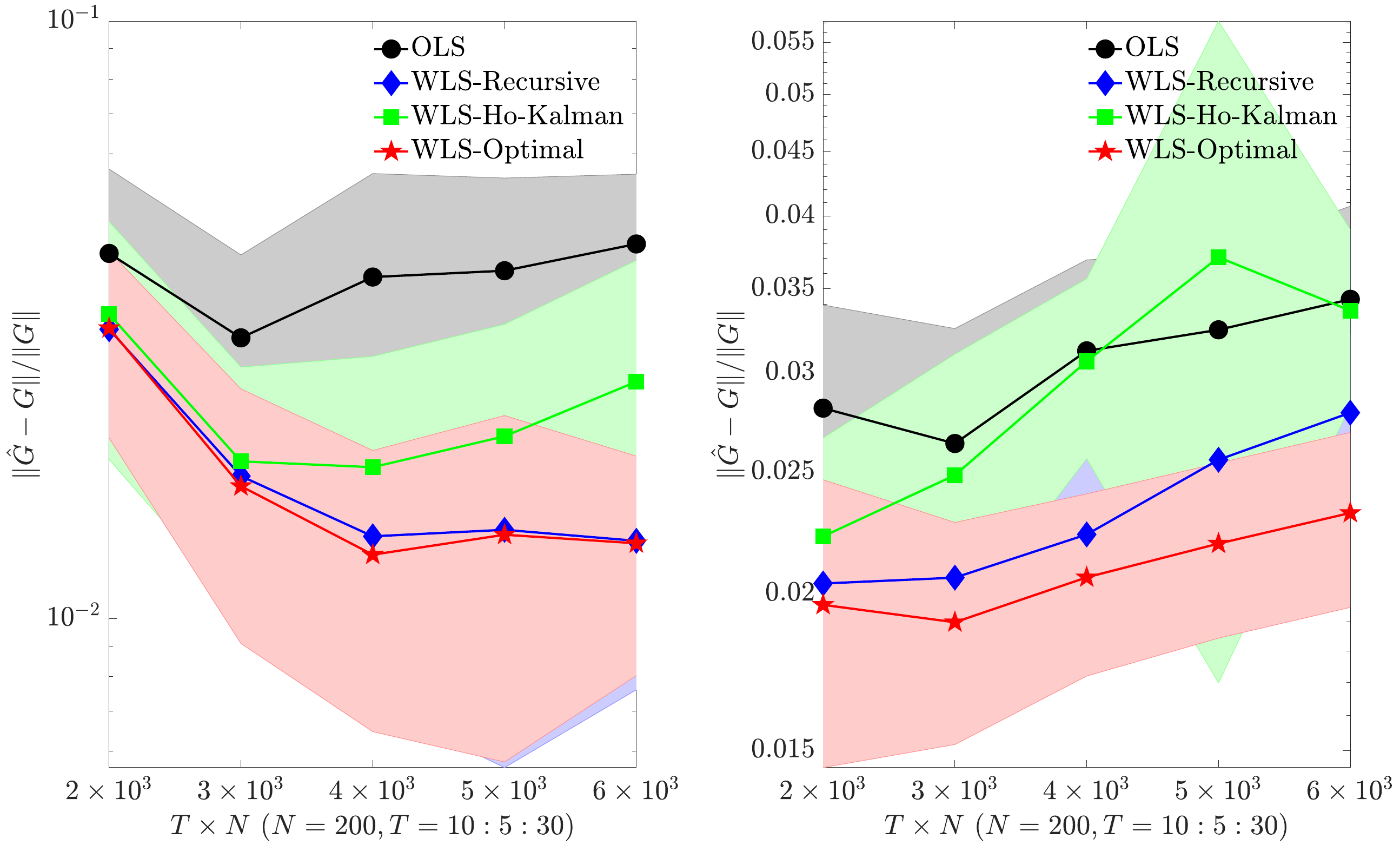}
	\caption{Estimation errors of SISO (left) and MIMO (right).}
	\label{F2}
\end{figure}

\vspace{-1.5mm}
Here we compare WLS with OLS using numerical examples. Since it has been showed that the method in \cite{zheng2020non} has better performance than OLS in \cite{simchowitz2019learning} and \cite{sun2020finite}, we mainly compare WLS with the one proposed in \cite{zheng2020non}. Two numerical examples are used, one is a marginally stable SISO system, where \begin{equation*} \label{E46}
	A = \begin{bmatrix}1&0.2 \\ 0&1\end{bmatrix}, B = \begin{bmatrix}0\\ 1\end{bmatrix},  C = \begin{bmatrix}1\\ 0\end{bmatrix}, D = 0,
\end{equation*}
and $K$ is randomly generated by MATLAB function $\rm{randi([-2,2],2,1)}$, and the other is a stable MIMO system, where
\begin{equation*} \label{E47}
	\begin{split}
		{A} &= \begin{bmatrix}
			0.67&0.67&0&0 \\ 
			-0.67&0.67&0&0 \\ 
			0&0&-0.67&-0.67 \\
			0&0&0.67&-0.67\end{bmatrix}, 
		{B} = \begin{bmatrix}
			0.65&-0.52 \\ 
			1.96&0.48 \\ 
			4.31&-0.48 \\
			-2.64&-0.34\end{bmatrix}, \\
		{C} &= \begin{bmatrix}
			-0.37&0.07&-0.52&0.58 \\ 
			-0.89&0.75&0.11&0.09\end{bmatrix},
		{K} = \begin{bmatrix}
			-0.69&-0.14 \\ 
			0.17&0.56 \\ 
			0.64&-0.46 \\
			-0.94&0.10 \end{bmatrix},	
	\end{split}
	\nonumber
\end{equation*}
and $ D = 0	$. We choose the variance of inputs and noises $\sigma_u = \sigma_e = 1$ in our simulations. Two types of experiments are implemented, one is fixing the length $T=10$, and varying the number of multi-rollouts $N=50:50:500$, and the other is fixing $N=200$, and varying the length $T=10:5:30$. For each case, we run 50 Monte Carlo trials. As for the evaluation, the estimation error in the Markov parameters $G$ is defined as $\lVert \hat G - G\rVert / \lVert G\rVert$. The results of those two setups are shown in Figures \ref{F1} and \ref{F2}, where the solid lines mean average errors of each method in 50 Monte Carlo trials, and the colored regions around them mean the variances in 50 trials. 

As shown in Figure \ref{F1}, for both SISO and MIMO systems, when the length $T$ is fixed, the estimation errors of those least-squares algorithms decrease when the number of rollouts $N$ increases. In terms of the comparison between OLS and WLS, with the optimal weighting matrices, WLS performs better than OLS. For WLS with estimated weighting matrices, with the increase of the number of samples, both of the two estimation algorithms are better than OLS. Furthermore, the recursive computing algorithm is better than the Ho-Kalman algorithm, and it approaches the optimal WLS closely.

As shown in Figure \ref{F2}, it is interesting to see that varying the length $T$ has different impacts on SISO and MIMO systems, especially for WLS. For OLS, when $N$ is fixed, increasing $T$ seems to have little impact on the normalized errors of the Markov parameters of SISO and MIMO systems, which is consistent with the observation in \cite{zheng2020non}. For WLS, the effects are not uniform. For the SISO system, it seems that increasing $T$ reduces the estimation error, while for the MIMO system, it is clear that increasing $T$ increases the estimation error. All these observations indicate that our bound is not tight. It is interesting to derive a tighter bound in the future.
\vspace{-1.5mm}
\section{Conclusion} \label{Sct6}
\vspace{-1.5mm}
In this paper, WLS is applied to identify Markov parameters of LTI systems using multiple trajectories. We show that for WLS, the bound on the estimation error is tighter compared to OLS. Furthermore, we provide two methods to consistently estimate the optimal weighting matrix, and convergence rates of these estimates are also derived. In the future, it will be interesting to incorporate this method in the framework of subspace identification to improve its performance. Also, this method will be extended to identify the Markov parameters of stable LTI systems using a singe trajectory. 
\vspace{-2.5mm}
\bibliography{ifacconf}             
                                                   







\end{document}